\documentclass[aps,pra,twocolumn,floats,superscriptaddress,tighten,eqsecnum]{revtex4}
\usepackage{amssymb}
\usepackage{amsbsy}
\usepackage{amsmath}
\usepackage{graphicx}
\usepackage{graphics}
\usepackage{setspace}
\usepackage{array}
\usepackage{color}
\usepackage{fontenc}
\usepackage{textcomp}
\usepackage{rotating}
\usepackage{bm}

\DeclareMathOperator{\diag}{diag}

\DeclareMathOperator{\Tr}{Tr}
\DeclareMathOperator{\sgn}{sgn}

\newcommand{\kc}{\msf{k}}
\newcommand{\kb}{\bar{\msf{k}}}
\newcommand{\e}{\varepsilon}

\newcommand{\vex}[1]{\bm{\mathrm{#1}}}

\newcommand{\pup}[1]{{\scriptscriptstyle{({#1})}}}

\newcommand{\pupsf}[1]{{\scriptscriptstyle{(\mathsf{{#1}})}}}

\newcommand{\msf}[1]{\mathsf{#1}}
\newcommand{\ket}[1]{| {#1} \rangle}
\newcommand{\bra}[1]{\langle {#1} |}
\newcommand{\braless}[1]{\left\langle {#1} \right.}
\newcommand{\tauh}{\hat{\tau}}
\newcommand{\kaph}{\hat{\kappa}}
\newcommand{\sigh}{\hat{\sigma}}

\newcommand{\Sh}{\hat{S}}
\newcommand{\Sb}{\hat{\bm{S}}}
\newcommand{\T}{\mathsf{T}}
\newcommand{\Mp}{\hat{M}_{\mathsf{P}}}
\newcommand{\Mps}{\hat{M}^{\pupsf{S}}_{\mathsf{P}}}
\newcommand{\Mt}{\hat{M}_{\mathsf{T}}}

\newcommand{\hs}{\hat{h}_{\scriptscriptstyle{\mathsf{S}}}}

\newcommand{\hsP}{\hat{h}^{\pup{P}}_{\scriptscriptstyle{\mathsf{S}}}}
\newcommand{\Hs}{H_0^\pupsf{S}}
\newcommand{\His}{H_I^\pupsf{S}}
\newcommand{\ord}[1]{\bm{\mathit{O}}\left(#1\right)}
\newcommand{\parr}{\partial}
\newcommand{\parb}{\bar{\partial}}
\newcommand{\Pc}{\mathsf{P}}
\newcommand{\Pb}{\bar{\mathsf{P}}}

\newcommand{\bsub}{\begin{subequations}}
\newcommand{\esub}{\end{subequations}}
\newcommand{\ts}[1]{{\textstyle{{#1}}}}

\begin{document}
\title{Disorder-enhanced topological protection and universal quantum criticality in a spin-3/2 topological superconductor}
\author{Sayed Ali Akbar Ghorashi}
\affiliation{Texas Center for Superconductivity and Department of Physics, University of Houston, Houston, Texas 77204, USA}
\author{Seth Davis}
\affiliation{Department of Physics and Astronomy, Rice University, Houston, Texas 77005, USA}
\author{Matthew S.\ Foster}
\affiliation{Department of Physics and Astronomy, Rice University, Houston, Texas 77005, USA}
\affiliation{Rice Center for Quantum Materials, Rice University, Houston, Texas 77005, USA}
\date{\today\\}

\newcommand{\be}{\begin{equation}}
\newcommand{\ee}{\end{equation}}
\newcommand{\bea}{\begin{eqnarray}}
\newcommand{\eea}{\end{eqnarray}}
\newcommand{\h}{\hspace{0.30 cm}}
\newcommand{\vs}{\vspace{0.30 cm}}
\newcommand{\n}{\nonumber}
\begin{abstract}	
We study the Majorana surface states of higher-spin topological superconductors (TSCs)
that could be realized in ultracold atomic systems or doped semimetals with spin-orbit coupling.  
As a paradigmatic example, we consider a model with $p$-wave pairing of spin-3/2 fermions that generalizes ${}^3\text{He-}B$. 
This model has coexisting linear and cubic dispersing Majorana surface bands. 
We show that these are unstable to interactions, which can generate a spontaneous surface thermal quantum Hall effect (TQHE). 
By contrast, nonmagnetic quenched disorder induces a surface conformal field theory (CFT) that is stable against weak interactions:
topological protection is \emph{enhanced} by disorder. 
Gapless surface states of higher-spin TSCs could therefore be robustly realized in solid state systems, where disorder is inevitable. 
The surface CFT is characterized by universal signatures that depend only on the bulk topological
winding number, and include power-law scaling of the density of states, a universal multifractal spectrum
of local density of states fluctuations, and a quantized ratio of the longitudinal thermal conductivity $\kappa_{xx}$ divided by temperature $T$. 
By contrast, $\kappa_{xx}/T$ for the clean surface without TQHE order would diverge as $T \rightarrow 0$. 
Since disorder stabilizes the conducting Majorana surface fluid and quantizes thermal transport, 
our results suggest a close analogy between bulk TSCs and the integer quantum Hall effect.
\end{abstract}
\maketitle


\section{Introduction}

The precise quantization of the Hall conductance is the heart of the integer quantum Hall effect (IQHE).
The quantization condition holds irrespective of material parameters, quenched disorder, 
or residual electron-electron interactions. The robustness of the IQHE is due to the tight entwining 
of the bulk topology (most easily defined for a clean system), Anderson localization (due to quenched disorder, 
always required in practice to resolve the plateaux), and the chiral edge states whose flow can never be degraded 
by perturbations such as disorder or interactions, but only redirected. 
Topological protection in the IQHE locks an easily measured observable directly to a bulk winding number.

	\begin{figure}[b!]
	\includegraphics[angle=0,width=.42\textwidth]{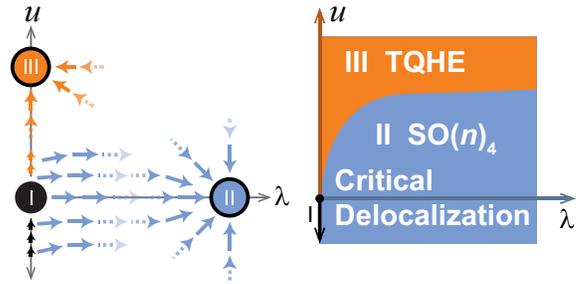}
	\caption{Schematic renormalization group (RG, left) and parameter (right)
	phase diagram: 2D Majorana fermion surface fluid of a bulk spin-3/2 topological superconductor
	with $p$-wave pairing and winding number $\nu = 4$.
	The axes denote surface perturbations: 
	the interaction strength $u$ and quenched disorder strength $\lambda$. 
	In the absence of disorder ($\lambda = 0$), the clean surface (``I'') is marginally unstable to 
	spontaneous time-reversal symmetry breaking 
	[thermal quantum Hall effect (TQHE) order, (``III'')] for $u > 0$. 
	Disorder $\lambda > 0$ is a strong perturbation that drives the surface
	into a critically delocalized, time-reversal symmetric state [SO$(n)_4$ CFT,
	``II'']. The latter is stable against weak interaction effects \cite{WZWP4}.
	Disorder is formally irrelevant to III because it can be viewed as
	a gapped, paired Majorana BCS condensate. 
	RG results near I and II are obtained by analytical and numerical calculations, 
	while III is confirmed by mean-field theory.
	The boundary between II and III on the right likely indicates
	a first order transition, but neither its nature nor its precise shape 
	in the $\lambda$-$u$ plane is determined here. 
	The surface thermal conductivity is precisely quantized in \emph{both} 
	phases II and III:
	$\left\{\kappa_{xx},\kappa_{xy}\right\} = \{4/\pi,0\} \, \kappa_\circ$ in II \cite{SRFL2008,WZWP3}
	and
	$\left\{\kappa_{xx},\kappa_{xy}\right\} = \{0,\pm 2\} \, \kappa_\circ$ in III \cite{KaneFisher,ReadGreen2000,Capelli02,Zhang2011,Ryu2012,Stone2012}.
	Here $\kappa_\circ \equiv \pi^2 k_B^2 T/6 h$.
	}
	\label{Fig--PD}
	\end{figure}

In the last decade there has been an explosion of interest in new forms of topological matter, 
driven by the 
discoveries of topological insulators and gapless
topological phases \cite{TIRev2010,BernevigHughes2013,TSCRev2016,TSCRev2016B}. 
Despite this progress, a three-dimensional analog
of the IQHE that ties a robust surface transport signature directly to a bulk winding number
remains lacking. One potentially promising route is to look for a generalization of Helium 3B (${}^3\text{He-}B$), 
the only known bulk topological superfluid predicted to host
a gapless surface fluid of unpaired Majorana fermions \cite{Volovik2003,SRFL2008}. 

Previous theoretical work \cite{SRFL2008,WZWP2,WZWP4,YZCP1,WZWP3}
has shown that the Majorana surface fluid of a model
spin-1/2 bulk topological superconductor (TSC) can be robust to both disorder and
interaction effects, and should exhibit a universal surface thermal conductivity
proportional to the bulk winding number \cite{WZWP3}. 
In all of these works, the form of the 2D surface theory was always assumed to be relativistic,
with $|\nu|$ ``colors'' of linearly-dispersing Majorana fermions coupled via
interactions and/or quenched disorder; $\nu$ denotes the winding number. 
A key unanswered question is whether the physics (e.g., universal thermal conductivity) is tied to this 
simplifying assumption, or instead represents a robust aspect of generic bulk TSCs. 

Recent theoretical work has turned to higher-spin TSCs, with potential applications
to 
alkaline and alkaline-earth
ultracold atoms \cite{Wu2016} or doped semimetals with spin-orbit coupling
\cite{Fang2015,Paglione2016,Brydon2016}. 
In this
paper, 
we consider the surface states of a spin-3/2 generalization of ${}^3\text{He-}B$ with isotropic
$p$-wave pairing \cite{Wu2016}.
A novel feature is that the surface Majorana fluid exhibits coexisting linear and cubic 
bands.
Cubic surface bands were also predicted 
in a closely related model \cite{Fang2015}
that may be relevant for superconducting half-Heusler alloys.
Due to the van Hove singularity, one might expect that any residual interactions between surface Majorana particles
would produce a strong instability. Surprisingly, we show that interactions are only \textit{marginally relevant}:
only attractive interactions induce spontaneous time-reversal symmetry breaking and lead to a
surface thermal quantum Hall effect (TQHE) \cite{SRFL2008,WZWP2}. This weak instability is tied to the 
strong constraints imparted by Pauli exclusion to a Majorana gas, despite the density of states divergence.
Repulsive interactions are marginally irrelevant; their main effect would be to 
generate a finite longitudinal surface thermal conductivity $\kappa_{xx}$ at temperature $T > 0$ due to inelastic scattering. 
In the absence of impurity scattering the ratio $\kappa_{xx}/T$ would diverge as $T \rightarrow 0$.

By contrast, 
nonmagnetic
quenched disorder proves to be a strong perturbation.
Using exact diagonalization to study the noninteracting dirty surface, we show
that disorder induces scaling consistent with
a critical, exactly solvable conformal field theory (CFT) SO$(n)_4$ \cite{WZWP4}.
(Here $n \rightarrow 0$ denotes the number of replicas.) 
The CFT governs the divergence of the global density of states and the statistics of 
the single-particle wave functions. 
The level 
of the current algebra (=4)
is also the modulus of the bulk winding number $|\nu|$ for our $p$-wave model. 
This is the same result that obtains for spin-1/2 models of TSC surface states
studied previously \cite{WZWP4,WZWP3}. 
In the spin-1/2 case with winding number $\nu$, the clean surface fluid is a free fermion (level one) CFT due
to the relativistic dispersion.
The emergence of another CFT with level $|\nu|$ 
in the presence of disorder follows from certain rules in these theories 
(conformal embeddings \cite{WZWP4,WZWP3}).

Here the situation is very different.
The clean Majorana fluid of the spin-3/2 model is not a CFT, as evidenced by the cubic
dispersion. Moreover, a standard derivation of the effective surface theory with disorder
would incorrectly predict a thermal metal with weak antilocalization \cite{SenthilFisher2000,WZWP4}.
Properties of this metal would depend on the bare disorder strength and would vary slowly with system size.  
Our numerics instead show universal scaling that is independent of the disorder strength.
In conjunction with the conformal embedding argument for the spin-1/2 case, the results obtained here
empirically suggest a deep relation between the topology of the bulk
and the CFT describing the \emph{disordered surface} of a TSC, despite
the fact that the clean surface theories can fundamentally differ. 
Technically, it means that the topology precisely tunes the surface field theory to the 
conformal fixed point made possible by a Wess-Zumino-Novikov-Witten term \cite{WZWP4}. 
Without this fine-tuning, this fixed point is \emph{unstable} to the thermal metal phase 
(despite the WZNW term) \cite{WZWP3}. 

The SO$(n)_4$ CFT that describes the disordered Majorana surface fluid  
is known to be 
protected
against weak interaction effects \cite{WZWP4}.
We conclude that disorder \emph{stabilizes} the surface Majorana fluid for this spin-3/2
model, and this implies that higher-spin TSCs could be robustly protected. 
The thermal Hall conductivity $\kappa_{xy}$ divided by temperature $T$ 
of the surface TQHE is quantized and universal \cite{KaneFisher,ReadGreen2000,Capelli02,Zhang2011,Ryu2012,Stone2012}:
$\kappa_{x y} = W \, \kappa_\circ$, where $W \in \mathbb{Z}$ is the \emph{surface winding number}
and
\begin{align}\label{kapcirc}
	\kappa_\circ/T = \pi^2 k_B^2/6 h. 
\end{align}
What is more important here is that \cite{SRFL2008,WZWP3}
\begin{align}\label{QuantKappa}
	\lim_{T \rightarrow 0}
	\frac{\kappa_{xx}}{T} 
	= 
	\frac{|\nu|}{\pi}
	\frac{\kappa_\circ}{T},
	\quad
	\kappa_{x y}
	=
	0,
\end{align}
for the disorder-induced surface CFT (which preserves time-reversal symmetry).  
Here the winding number $\nu = 4$. 
Eq.~(\ref{QuantKappa}) implies that the low-temperature thermal conductivity is quantized by
the bulk winding number, independent of both disorder and interactions. 
Since disorder stabilizes the surface and induces a quantized thermal conductivity, 
bulk TSCs appear to be closely analogous to the integer quantum Hall effect in two dimensions. 	
Our results are summarized by the phase diagram in Fig.~\ref{Fig--PD}. 

This paper is organized as follows.
In Sec.~\ref{Sec: Model}, we define the bulk model and describe the form of the surface states.
We then summarize our results regarding the marginal instability of the clean surface,
and the universal quantum criticality of the disordered one. 
The rest of the paper explains key technical details. 
Sec.~\ref{Sec: Clean} shows the derivation of the surface state Hamiltonian
and the calculation of the surface winding number in the presence
of explicit time-reversal symmetry breaking.
The effects of interactions on the clean surface 
are treated using one-loop renormalization and mean field theory.
Sec.~\ref{Sec: Dirty} describes the incorporation of disorder,
and provides details of the numerical diagonalization scheme.


\section{Model and results \label{Sec: Model}}

\subsection{Bulk and surface models}

We consider a system of spin-3/2 fermions. In the absence
of pairing, we assume a bulk Hamiltonian of the form
\begin{align}\label{HLutt}
	H_0 
	=
	\int \frac{d^3 \vex{k}}{(2 \pi)^3}
	\,
	c^\dagger(\vex{k})
	\,
	\left[
	k^2 
	- 
	\gamma 
	\left(\Sb\cdot\vex{k}\right)^2
	\right]
	\,
	c(\vex{k}),
\end{align}
where $\vex{k} = \{k_{x,y,z}\}$. 
The 4-component fermion field 
$c(\vex{k})\rightarrow c_{m_s}(\vex{k})$ has 
$\Sh^z$-label
$m_s \in \{\frac{3}{2},\frac{1}{2},-\frac{1}{2},-\frac{3}{2}\}$;
$\Sb=\{\Sh^{x,y,z}\}$ are spin-3/2 operators.
Eq.~(\ref{HLutt}) is an isotropic version of the
Luttinger Hamiltonian \cite{YuCardona} used to
model heavy and light hole bands in zinc-blende semiconductors;
the parameter $\gamma$ measures the strength of effective spin-orbit coupling (SOC)
amongst the states of the 3/2 multiplet. 
Here we have set $2 m = 1$ in the first term ($m$ is the band mass
in the absence of SOC).  
We assume that $\gamma < 4/9$, so that both
bands ``bend up.'' The situation where bands bend oppositely 
is relevant for half-Heusler alloys; in that case similar
Majorana surface states can arise with bulk $p$-wave pairing \cite{Fang2015},
but the winding numbers differ \cite{Ghorashi2017}. 

We assume isotropic $p$-wave pairing of spin-3/2 fermions \cite{Wu2016}: 
\begin{align}\label{HDef}
	H
	=
	\frac{1}{2}
	\int \frac{d^3 \vex{k}}{(2 \pi)^3}
	\,
	\chi^\dagger(\vex{k})
	\,
	\hat{h}(\vex{k})
	\,
	\chi(\vex{k}),
\end{align} 
where the $8\times8$ Bogoliubov-de Gennes Hamiltonian is 
\begin{align}\label{hBdG}
	\hat{h}(\vex{k})
	=
	\left[
	k^2 
	- 
	\mu
	-
	\gamma 
	\left(\Sb\cdot\vex{k}\right)^2
	\right]
	\sigh^3
	+
	\Delta_p
	\left(
	\Sb
	\cdot
	\vex{k}
	\right)
	\sigh^2.
\end{align}
Here $\mu$ is the chemical potential and $\Delta_p$ the BCS gap parameter. 
The 8-component field in Eq.~(\ref{HDef}) 
has the particle-hole space decomposition 
\begin{align}\label{chiDef}
	\chi
	(\vex{k})
	\equiv
	\begin{bmatrix}
	c(\vex{k})	\\
	(-i \hat{R}) \, \left[c^\dagger(-\vex{k})\right]^\T 
	\end{bmatrix},
\end{align}
where $\T$ denotes the transpose in spin-3/2 space. 
The Pauli matrices $\sigh^{1,2,3}$ in Eq.~(\ref{hBdG}) act on particle-hole space.
In Eq.~(\ref{chiDef}), 
$\hat{R}$ is an antisymmetric 4$\times$4 matrix satisfying
\begin{align}\label{RDef}
	\hat{R} \, (\Sb)^\T \, \hat{R}^{-1}= - \Sb, \quad \hat{R}^2 = -\hat{1}. 
\end{align}

The field $\chi$ satisfies the ``Majorana'' condition
$\chi^\dagger(\vex{k}) = i \, \chi^\T(-\vex{k}) \, \Mp$,
where
$\Mp = -i  \sigh^2 \, \hat{R} = \Mp^\T$.
This implies the \emph{automatic}
particle-hole symmetry
$
	- \Mp^{-1} \, \hat{h}^\T(-\vex{k}) \, \Mp = \hat{h}(\vex{k}). 
$
Time-reversal invariance is encoded as 
$
	\Mt^{-1} \, \hat{h}^\T(-\vex{k}) \, \Mt = \hat{h}(\vex{k}),
$
where $\Mt = - \Mt^\T = \sigh^3 \, \hat{R}$. 
Combining particle-hole and time-reversal \cite{Wu2016,SRFL2008}
gives the effective chiral condition
$
	- \sigh^1 \, \hat{h}(\vex{k}) \, \sigh^1 = \hat{h}(\vex{k}).
$
With all of these symmetries the model belongs to class DIII \cite{SRFL2008}.
The bulk winding number is $\nu = 4$ \cite{Wu2016} so long as $\mu > 0$ and $\gamma < 4/9$. 	

The effective surface Hamiltonian obtains by terminating the 
system in the $z$-direction and diagonalizing $\hat{h}(k_x,k_y,-i \partial_z)$.
The momentum $\vex{k} = \{k_{x,y}\}$ labels propagation parallel to the surface. 
For $\vex{k} = 0$ and hard wall boundary conditions, there are four zero energy bound states
$\{\ket{\psi_{0,m_s}}\}$. 
The most important feature is that the particle-hole ``spin'' locks
to the physical spin at the surface: 
\begin{align}\label{psi_0}
	\ket{\psi_{0,m_s}} 
	=
	\ket{\sigma^1 = \sgn(m_s)}
	\otimes
	\ket{m_s}
	\otimes
	\ket{f_{m_s}}.
\end{align}
The particle-hole spin points along the $+ \sigma^1$ ($-\sigma^1$) direction
for positive (negative) $m_s$.
In Eq.~(\ref{psi_0}), $\braless{z \!}\ket{f_{m_s}} = f_{m_s}(z)$ denotes the bound state envelope function. 

Since time-reversal gives the effective chiral condition 
$- \sigh^1 \, \hat{h}(\vex{k}) \, \sigh^1 = \hat{h}(\vex{k})$ in the bulk,
the locking condition implies that the effective surface Hamiltonian $\hs$
satisfies 
\begin{align}\label{Chiral-Surf}
	- \tauh^3 \, \hs(\vex{k}) \, \tauh^3 = \hs(\vex{k}),
\end{align}
where $\tauh^3 = +1$ ($-1$) for $m_s > 0$ ($m_s < 0$). 
We introduce two mutually commuting species of Pauli matrices: 
$\{\tauh^{1,2}\}$ anticommute with $\tauh^3$ and act on the 
$\sgn(m_s)$ space, 
while 
$\{\kaph^{1,2,3}\}$ mix the $3/2$ and $1/2$ states with the same 
sign.
E.g., 
$
	\tauh^3 
	=
	\diag(1,1,-1,-1),
$
$
	\kaph^3 
	=
	\diag(1,-1,1,-1). 
$
The matrix $\hat{R} = i \tauh^1 \kaph^2$. 
Then the locking condition implies the
\emph{automatic}
surface particle-hole symmetry
\begin{align}\label{PH-Surf}
	- 
	\Mps
	\,
	\hs^\T(-\vex{k})
	\,
	\Mps
	= 
	\hs(\vex{k}), 
	\;\;
	\Mps = \tauh^2 \kaph^2.
\end{align}

The form of $\hs(\vex{k})$ is constrained by Eqs.~(\ref{Chiral-Surf}) 
and (\ref{PH-Surf}), as well as rotational invariance in the plane. 
An explicit  k$\cdot$p calculation 
gives the low-energy form \cite{Wu2016}
\begin{align}\label{hsDef}
	\hs(\vex{k}) 
	=&\,
	{\ts{\frac{i}{4}}}
	\left(
	\tauh^+ \kaph^-
	\,
	\kc
	-
	\tauh^- \kaph^+
	\,
	\kb
	\right)
	+
	{\ts{\frac{c}{2}}}
	\left(
	\tauh^+
	\,
	\kc^2 
	+
	\tauh^-
	\kb^{2}
	\right)
\nonumber\\
	=&\,
\begin{bmatrix}
	0 & 0 & c \, \kc^2 & 0 \\
	0 & 0 & i \, \kc & c \, \kc^2 \\
	c \, \kb^2 & -i \, \kb & 0 & 0 \\
	0 & c \, \kb^2 & 0 & 0
	\end{bmatrix}.	
\end{align}
Here, 
$\{\kc,\kb\} = k_x \mp i k_y$, 
$\tauh^{\pm}=\tauh^1 \pm i \tauh^2$,
$\kaph^{\pm}=\kaph^1 \pm i \kaph^2$,
the coefficient for the linear dispersion is normalized to one,
and $c$ is a real constant with units of length.
For weak SOC ($0 < \gamma \ll 1$), 
it is easy to see that $c \propto \gamma / \Delta_p$. 
Without SOC, second-order k$\cdot$p theory gives
$c = 0$ so that the $m_s = \pm 3/2$ bands remain flat.
Nonzero $c$ is symmetry-allowed and thus expected in the 
generic situation; an alternative route incorporates additional 
small $s$-wave pairing.
In Eq.~(\ref{hsDef}), we neglect terms cubic in $k_{x,y}$
because these do not modify the low-energy dispersion relations. 

In the limit $k \rightarrow 0$, Eq.~(\ref{hsDef}) exhibits 
linear and cubic bands \cite{Wu2016}:
\begin{align}\label{EBands}
\begin{aligned}
	\e_{1}(k) =&\, \frac{k}{2}\left(\sqrt{1 + 4 c^2 k^2} + 1\right) \simeq k + \ord{k}^3,
\\
	\e_{3}(k) =&\, \frac{k}{2}\left(\sqrt{1 + 4 c^2 k^2} - 1\right) \simeq c^2 k^3 + \ord{k}^5.
\end{aligned}
\end{align}	
Both bands become quadratic at large $k$. 
Due to the cubic band, the system has a van Hove singularity
in the density of states $\nu(\e \rightarrow 0) \sim \e^{-1/3}$.


\subsection{Surface perturbations}

The Hamiltonian 
for the surface fermion fluid is given by 
\begin{align}
	\Hs = \frac{1}{2} \int d^2 \vex{r} \, \eta^\T \, \Mps \, \hs \, \eta,
\end{align}
where $\eta \rightarrow \eta_{m_s}$ is a four-component Majorana spinor and $\vex{r}$ is 
the position vector. 
Local bilinear (``potential'') perturbations must obey surface particle-hole
symmetry. There are 6 Hermitian terms without derivatives of the form
$\frac{1}{2} \int d^2 \vex{r} \, \eta^\T \, \Mps \, \hat{\Lambda} \, \eta$,
where $\Lambda \in \{\tauh^{1,2,3}\}$ or $\{\kaph^{1,2,3}\}$ satisfies Eq.~(\ref{PH-Surf}). 
These are classified by symmetry. 

Under planar rotations, $\tauh^3$ and $\kaph^3$ are scalars, $\kaph^{1,2}$ transform
like a vector, and $\tauh^{1,2}$ transform like second-rank tensor components. 
Only $\tauh^{1,2}$ are time-reversal even [Eq.~(\ref{Chiral-Surf})]; the rest are
odd. 

\begin{figure}[b!]
	\includegraphics[angle=0,width=.2\textwidth]{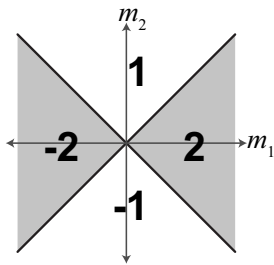}
	\caption{Phase diagram of the clean spin-3/2 Majorana surface fluid 
	in the presence of time-reversal symmetry-breaking mass terms
	$m_1$ and $m_2$. All states are TQHE plateaux with winding number
	as indicated, computed via the surface Green's function \cite{Volovik2003}. 
	The combination $m_2 = m_1/2$ corresponds to the spin-3/2 
	operator $\hat{S}^z$, as could be introduced via Zeeman coupling
	to an external magnetic field.
	The conditions $m_1 = \pm m_2$ are gapless plateau transitions.
	}
	\label{Fig--TQHEP}
	\end{figure}

A generic combination $\hat{\Lambda} = m_1 \, \tauh^3 + m_2 \, \kaph^3$ 
breaks time-reversal and induces a 
gapped
surface thermal quantum Hall (TQH) state 
\cite{ReadGreen2000,TSCRev2016}. We compute the \emph{surface winding number} $W$ 
using the Green's function \cite{Volovik2003}; the result is shown
in Fig.~\ref{Fig--TQHEP}. The lines $m_1 = \pm m_2$ are gapless surface plateau transitions.
The maximum winding number and surface gap for fixed $\sqrt{m_1^2 + m_2^2}$ is achieved for  
$|m_1| > |m_2|$, i.e.\ $\tauh^3$ order. 
Finally, we note that the spin operator $\hat{S}^z$ corresponds to $m_1 = 2 m_2$,
so that an external Zeeman field would induce the $W = \pm 2$ plateaux.

\subsection{Marginal instability of the clean surface}

Although we treat the gapped
bulk as an effectively non-interacting mean field Hamiltonian, we must consider
the effects of residual interactions on the surface Majorana fluid carefully. This is 
because the latter is gapless and exhibits a diverging density of states. 
In a superconductor, interactions at the surface are expected to be
short-ranged due to screening by the bulk. These residual interactions 
can be mediated by virtual fluctuations of the ``massive'' electromagnetic field.
Here we posit the form of the interactions based on symmetry and Pauli exclusion.

Because $\eta$ is a four-component Majorana field, there is only a single interaction 
term without derivatives that we can write; others with derivatives are less relevant.
Labeling the components as $\eta \rightarrow \eta_{1,2,3,4}$, 
\begin{align}\label{HisDef}
	\His
	\equiv&\,
	u
	\int d^2 \vex{r}
	\,
	\eta_{1}
	\eta_{2}
	\eta_{3}
	\eta_{4}
\nonumber\\
	=&\,
	\mp
	\frac{u}{8}
	\int d^2 \vex{r} \, (\eta^\T \, \Mps \, \hat{\Lambda} \, \eta )^2,
\end{align}
where the minus (plus) sign corresponds to $\hat{\Lambda} = \tauh^3$ ($\kaph^3$). 
Thus $u > 0$ is an attractive (repulsive) interaction in the $\tauh^3$ ($\kaph^3$) channel. 
The coupling $u$ has units of length. 
The sign of $u$ could be determined by integrating out the bulk superfluid and the electromagnetic
field, but we will not do so here.

Given that the noninteracting surface fluid has coexisting linear and cubic bands,
it is not a priori obvious how to assess the relevance of $u$ from a renormalization
group (RG) perspective. Moreover, the van Hove singularity suggests that nonzero $u$ will
induce bad infrared behavior. In fact one-loop perturbation theory gives the
simple vertex correction,
\begin{align}\label{Vertex}
	\Gamma^{\pup{4}}
	=
	-
	\left[
	u
	+
	\left(u^2/4 \pi c\right)
	\ln(4 c \Lambda)
	\right],
\end{align}
where $\Lambda$ is the ultraviolet momentum cutoff. 
The correction is only logarithmic, and is cut in the infrared by the length scale $c$. 
This immediately implies the beta function,
\begin{align}\label{uBetaClean}
	\frac{d \tilde{u}}{d l}
	=
	\frac{\tilde{u}^2}{4\pi}
	+
	\ord{\tilde{u}^3},
\end{align}
where $\tilde{u} = u / c$ is the dimensionless coupling. 

The absence of bad infrared behavior in Eq.~(\ref{Vertex}) 
and the weakness of the ultraviolet singularity is due to 
Pauli exclusion, i.e.\ the fact that both linear and cubic components of the
Majorana Green's function must appear simultaneously in the loop. 
Eq.~(\ref{uBetaClean}) implies that $u > 0$ is a marginally relevant perturbation. 
Eq.~(\ref{HisDef}) suggests a natural interpretation in terms of 
TQH order with surface winding number $W = \pm 2$. 

We can confirm this picture with a mean-field calculation.
We decouple the interaction in Eq.~(\ref{HisDef}) with the 
order parameter $\mathcal{M} \equiv (u/2) \, \langle \eta^\T \, \Mps \, \tauh^3 \, \eta \rangle$. 
Zero-temperature mean-field theory gives 
\begin{align}\label{MFT}
	\mathcal{M}
	\simeq
	\frac{1}{c}
	\left(\frac{4.4}{2 \pi}\right)^3
	\frac{1}{\left[(2 \pi / \tilde{u}) - \ln(c \Lambda)\right]^{3}},
	\quad
	\tilde{u} \ll 1.
\end{align}

Physically we can associate nonzero $\mathcal{M}$ with \emph{surface} 
``i $s$'' (imaginary $s$-wave) pairing of the Majorana particles \cite{WZWP2,WZWP4,Nomura2015}.
This can be understood via the following argument.
In the bulk, one can write a local spin singlet, time-reversal odd pairing operator 
\begin{align}
	[-i c^\dagger \hat{R} \left(c^\dagger\right)^\T + \textrm{H.c.}]
	=
	\chi^\dagger \sigh^1 \chi.
\end{align}
Eq.~(\ref{psi_0}) implies that this bilinear projects
to $\eta^\T \Mps \tauh^3 \eta \propto \mathcal{M}$ at the surface. 
On the other hand, Eq.~(\ref{HisDef}) can also be written 
as proportional to 
$	
	- u \int d^2\vex{r} \, (\hat{S}^z)^2
$, 
implying surface magnetic order for
$u > 0$. Indeed, these disparate orders are unified in the surface fluid, due
to the strong spin-orbit coupling in the bulk and the locking condition [Eq.~(\ref{psi_0})].  
We might anticipate a generic order parameter of the form 
$\hat{\Lambda} = m_1 \, \tauh^3 + m_2 \, \kaph^3$ [c.f.\ Fig.~\ref{Fig--TQHEP}]
with $|m_1| > |m_2|$. Although we are confident that the surface resides in the
$W = + 2$  or $- 2$ plateau for $u > 0$, there are hints that mean-field theory fails
to correctly predict the admixture of $m_1$ and $m_2$.
For details, see Sec.~\ref{Sec: MFT}, below.


\subsection{Quenched disorder and universal surface quantum criticality}

In a solid state
realization, quenched disorder due to impurities and other defects is inevitable at the sample
surface. 
Now we consider the effects of disorder on the non-interacting surface states. 

We add real disorder potentials that couple to the time-reversal symmetric bilinear
perturbations $\tauh^{1,2}$ to Eq.~(\ref{hsDef}):
\begin{widetext}
\begin{align}\label{Dirtyhs}
	\hs
	\rightarrow&\,
	\hs
	+
	P_1(\vex{r}) \, \tauh^1 
	+
	P_2(\vex{r}) \, \tauh^2
	=
\begin{bmatrix}
	0 				& 0 				& c (-i \parr)^2 + \Pc(\vex{r}) 	& 0					\\
	0 				& 0				& \parr					& c (-i \parr)^2 + \Pc(\vex{r}) 	\\
	c (-i \parb)^2 + \Pb(\vex{r}) 	& -\parb			& 0					& 0					\\
	0				& c (-i \parb)^2 + \Pb(\vex{r})	& 0					& 0	
\end{bmatrix},			
\end{align}
\end{widetext}
where $\{\parr,\parb\} = \partial_x \mp i \partial_y$ and $\{\Pc,\Pb\} = P_1 \mp i P_2$. 
We consider only time-reversal
invariant disorder; equivalently, we require that there are no magnetic fields or
magnetic impurities at the surface.  
We assume Gaussian white noise  
disorder potentials $P_{1,2}(\vex{r})$ with common variance given by a 
dimensionless parameter $\lambda$. 
Due to the cubic dispersion, even weak disorder is expected to produce a strong effect.
Indeed, perturbative renormalization of $\lambda$ produces a quadratic infrared divergence,
and implies that $\lambda/c^2$ is the effective disorder strength. This has dimension 2
and thus corresponds to a strongly relevant perturbation of the clean surface band structure. 
To treat the disorder nonperturbatively, we diagonalize Eq.~(\ref{Dirtyhs})
 numerically. The calculation is performed in momentum space to avoid
fermion doubling issues \cite{YZCP1}. 
Details are described in Sec.~\ref{Sec: Num}, below.

	\begin{figure}[b!]
	\includegraphics[angle=0,width=.45\textwidth]{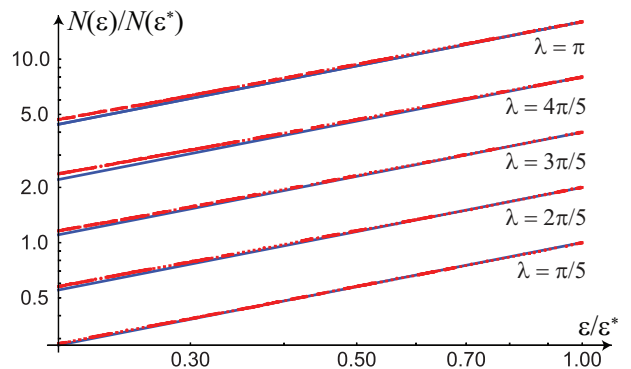}
	\caption{Numerical evidence for critical surface delocalization 
	in the presence of quenched disorder I: 
	Integrated local density of states (IDoS) $N(\e)$. 
	The exact prediction of the SO$(n)_4$ theory gives 
	$N(\e) \sim \e^{4/5}$ (blue solid lines).
	The clean theory has 
	$N(\e) \sim \e^{2/3}$ due to the
	van Hove singularity. 
	Data (red dotted lines) 
	is obtained from momentum-space exact diagonalization
	of the dirty surface Hamiltonian, without interactions.
	Results are presented for typical realizations of the disorder
	(i.e., there is no disorder-averaging). 
	Curves with different disorder strengths $\lambda$ 
	are labeled and shifted vertically for clarity.
	The system size consists of an 81 $\times$ 81 grid of momenta. 
	Irrespective of the nonzero disorder strength, 	
	the same critical scaling exponent is observed
	and is consistent with the SO$(n)_4$ theory.
	}
	\label{Fig--IDOS}
	\end{figure}

In spin-1/2 bulk TSCs with relativistic surface fluids, conformal embedding rules
establish certain 2+0-D conformal field theories (CFTs) as governing the properties of disordered,
noninteracting surface states \cite{WZWP4}. For a class DIII bulk with winding number 4,
that theory would predict the surface CFT SO$(n)_4$, where $n \rightarrow 0$ counts replicas. 
We will demonstrate that this theory also governs the dirty surface states of the spin-3/2 TSC.

The SO$(n)_4$ theory predicts 
\cite{WZWP4}
a diverging 
low-energy
global density of states (DoS) $\nu(\e) \sim \e^{-1/5}$.
Note that this is a weaker power law than the 1/3 van Hove singularity in the clean system. 
In Fig.~\ref{Fig--IDOS}, we compare numerical results for different disorder
strengths to the CFT prediction for the integrated DoS $N(\e) \equiv \int_0^\e d \e' \, \nu(\e')$.
We find good agreement irrespective of the disorder strength. 

\begin{figure}[t!]
	\includegraphics[angle=0,width=.45\textwidth]{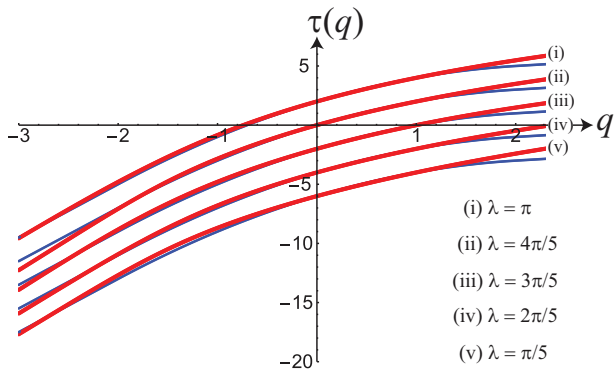}
	\caption{Numerical evidence for critical surface delocalization 
	in the presence of quenched disorder II: 
	Multifractal spectrum. 
	The exact prediction of the SO$(n)_4$ theory gives 
	Eq.~(\ref{tau(q)}) (blue).
	The clean theory would have $\tau(q) = 2(q - 1)$. 
	Data [red, labeled (i)--(v)] is obtained 
	as in Fig.~\ref{Fig--IDOS}. 
	The disorder strengths $\lambda$ are indicated for the numerical
	curves. 
	Curves with different disorder strengths are shifted vertically for clarity.
	The largest deviation occurs for $|q| > q_c$, where $q_c = 2$ is the
	multifractal termination threshold \cite{Chamon96,AndRev08,MFCP1}. 
	This is a finite resolution effect, since the slopes
	beyond $q_c$ are governed by the peaks and valleys of the wave function.
	}
	\label{Fig--MFC}
	\end{figure}

The disorder-induced spatial fluctuations of the critical surface wave functions 
are encoded in the multifractal spectrum $\tau(q)$. 
(For a recent review on multifractality at Anderson metal-insulator transitions, see e.g.\ Ref.~\cite{AndRev08}.)	
The SO$(n)_4$ theory predicts an exactly quadratic spectrum \cite{WZWP4}
for the low-energy wavefunctions,
\begin{align}\label{tau(q)}
	\tau(q)
	=
	(q-1)
	\left(2 - q / 2\right), 
	\qquad
	|q| \leq 2. 
\end{align}
Fig.~\ref{Fig--MFC} compares Eq.~(\ref{tau(q)}) to the numerical results.

Figs.~\ref{Fig--IDOS} and \ref{Fig--MFC} provide
strong evidence that the disordered, noninteracting spin-3/2 Majorana fluid
is governed by the SO$(n)_4$ theory. 
This is surprising because
a standard derivation \cite{SenthilFisher2000} of the disorder-induced effective field theory would predict
a thermal metal phase exhibiting weak antilocalization. 
Although the theory in \cite{SenthilFisher2000}
should be augmented by a Wess-Zumino-Novikov-Witten term
(see Sec.~\ref{Sec: DirtyTheory} for details),
this term does not alter the tendency towards antilocalization in the metallic phase \cite{WZWP4}. 
Because the clean density of states diverges for the surface Majorana fluid studied here, 
a diffusive metallic state would be generically expected. Yet this is inconsistent with our
numerical results [which instead match the SO$(n)_4$ CFT]. 
An important technical point is that the CFT is \emph{unstable} to 
the thermal metal phase [see Eq.~(\ref{DIIIBeta})]. 
It means that the CFT can only be realized if the system
is tuned to the SO$(n)_4$ fixed point. 
Our numerical results imply that this is exactly what happens. 

The same ``fine-tuning'' is required for the spin-1/2 TSCs. In that case,
however, there is a nonperturbative argument for it using conformal embedding
theory \cite{WZWP4}. Additional evidence in the spin-1/2 case obtains by 
comparing interaction (Altshuler-Aronov) corrections to transport via two
methodologies: (1) order-by-order in the interaction strength, in a \emph{fixed realization} of disorder,
and (2) within a disorder-averaged large winding number expansion. These give the same
result only if the disorder-averaged system is tuned to the CFT (in which case Altshuler-Aronov corrections vanish) 
\cite{WZWP3}. 
We do not have the conformal embedding dictionary \cite{WZWP4} utilized
for spin-1/2 TSCs, but our 
numerical 
results empirically suggest that there is an 
equivalence between the bulk topology and the CFT describing
the \emph{disordered surface} of a TSC \cite{WZWP4,YZCP1,WZWP3}, 
despite the fact that the clean surface theories can fundamentally differ.


\subsection{Stability, phase diagram, and quantized thermal conductivity}

We have seen that the clean surface
is marginally unstable to TQH order. 
We have also shown that quenched disorder is a strong perturbation that drives the 
non-interacting surface to a phase described by the SO$(n)_4$ CFT. 
It is known \cite{WZWP4} that interactions are strongly irrelevant to this CFT,
\begin{align}\label{uFlowCFT}
	\frac{d \tilde{u}}{d l} = -  \frac{\tilde{u}}{2} + \ord{\tilde{u}^2},	
\end{align}
where $\tilde{u}$ is the dimensionless coupling strength. 
Although multifractality can sometimes enhance interactions \cite{Feigelman07,Feigelman10,WZWP2}, 
that does not occur here. The reason is again Pauli exclusion: the interaction 
and second multifractal moment operators are distinct due to the complete antisymmetrization of the former \cite{WZWP4}. 
We therefore conclude that disorder stabilizes the surface Majorana fluid of this spin-3/2 TSC. 
This is our most important result. 

We note that Eq.~(\ref{uFlowCFT}) technically obtains from the \emph{dynamical version} of the 
SO$(n)_4$ theory. This is a 2+1-D theory of Majorana fermions propagating in space and time, 
whose disorder-averaged spatial correlations and dynamical scaling exponent are governed by 
the 2+0-D replicated CFT, see \cite{WZWP4} for details. Thus in Fig.~\ref{Fig--PD}, ``SO$(n)_4$'' really refers 
to this dynamical hybrid theory, which can also be expressed as a 
Wess-Zumino-Novikov-Witten Finkel'stein nonlinear sigma model 
(WZNW-FNLsM)	
\cite{WZWP4,WZWP3}.

The thermal conductivity of the WZNW-FNLsM receives no quantum interference corrections
due to disorder \cite{SRFL2008,WZWP3} at the conformal fixed point. 
Interaction-mediated Altshuler-Aronov corrections also vanish to at least order 1/$|\nu|$,
where $\nu$ is the bulk winding number. 
The absence of Friedel oscillations in any Majorana surface ``density'' implies
that these should be absent to all orders \cite{WZWP3}. 
Since the spin-3/2 Majorana surface fluid studied here realizes the SO$(n)_4$ theory
in the presence of disorder, we conclude that the ratio of the longitudinal
thermal conductivity and temperature is precisely quantized as $T \rightarrow 0$ 
[Eq.~(\ref{QuantKappa})].


\subsection{Summary}

In summary, we have derived surface states and surface effective Hamiltonian for a spin-3/2 
time-reversal invariant topological superconductor that hosts a cubic dispersion coexisting with 
the conventional linear Majorana cone. We have shown that in the clean limit, unlike the spin-1/2 case \cite{Maciejko15}, 
interactions are marginally relevant and lead to a BCS-type instability that gaps out the surface
and induces a thermal quantum Hall effect (TQHE) plateau. 
By contrast, quenched disorder gives the SO$(n)_4$ theory previously predicted for a spin-1/2
TSC with winding number $4$; this theory is stable to interaction effects. 
We conclude that disorder enhances topological protection. 
In the low-temperature limit, the ratio of the longitudinal thermal conductivity 
to temperature is predicted to be quantized and proportional to the bulk winding number,
as shown in Eq.~(\ref{QuantKappa}).

\begin{widetext}

\section{Ideal (clean) surface states and interactions \label{Sec: Clean}}

\subsection{Derivation of the surface Hamiltonian Eq.~(\ref{hsDef})}

\subsubsection{Luttinger Hamiltonian bulk}

To obtain the surface Hamiltonian, we consider a bulk superconductor in
the half space $z \geq 0$ with hard wall boundary conditions. 
We divide Eq.~(\ref{hBdG}) into two parts:
\bsub
\begin{align}
	\hat{h}
	=&\,
	\hat{h}_0(-i \partial_z)
	+
	\hat{h}_1(\vex{k},-i \partial_z),
\nonumber\\
\label{h0Def}
	\hat{h}_0
	=&\,
	\sigh^3
	\left( - \partial_z^2 - \mu \right)
	+
	\sigh^2
	\left[ 
	\Delta_p \Sh^z \left(-i \partial_z\right)
	+
	\Delta_s
	\right],
\\
\label{h1Def}
	\hat{h}_1
	=&\,
	\sigh^3
	\left\{	
	k^2
	-
	\gamma
		\left[
		\frac{1}{2}
		\left(
		\Sh^+ \kc 
		+
		\Sh^- \kb
		\right)		
		+
		\Sh^z
		(-i \partial_z)
		\right]^2
	\right\}
	+
	\sigh^2
	\frac{\Delta_p}{2}
	\left(
	\Sh^+ \kc 
	+
	\Sh^- \kb
	\right),
\end{align}
\esub
where $\vex{k} = \{k_{x,y}\}$ is the momentum parallel to the surface,
$\{\kc,\kb\} = k_x \mp i k_y$,
and
$\Sh^{\pm} = \Sh^x \pm i \Sh^y$ are spin-3/2 raising and lowering operators.
In Eq.~(\ref{h0Def}), we have added a time-reversal symmetric $s$-wave pairing
term proportional to $\Delta_s$. We will utilize this below in the case of
vanishing bulk spin-orbit coupling (SOC). Energies like $\Delta_s$ and $\mu$ 
have units of 1/(length)${}^2$, while $\Delta_p$ has units of 1/length.

In this subsection we will ignore $s$-wave pairing ($\Delta_s = 0$) and we will treat
the SOC term proportional to $\gamma$ as a small perturbation (although this is not necessary).
The surface eigenstates of $\hat{h}_0$ with zero transverse momentum satisfy 
$\hat{h}_0 \ket{\psi_{0,m_s}} = 0$ and take the form shown in Eq.~(\ref{psi_0}),
\begin{align}\label{psi0-ZM}
	\psi_{0,m_s}(z)
	=&\,
	f_{m_s}(z)
	\,
	\begin{bmatrix}
	1 \\ \sgn(m_s)
	\end{bmatrix}
	\ket{m_s},
\qquad
	f_{m_s}(z)
	=
	\frac{1}{\sqrt{\mathcal{N}_{m_s}}}
	\exp\left(-\frac{\Delta_p |m_s| z}{2}\right)
	\,
	\sin\left[z \sqrt{\mu - \frac{\Delta_p^2 m_s^2}{4}}\right].
\end{align}
Here the explicit 2-component spinor resides in particle-hole ($\sigma$) space; 
the four zero energy states are distinguished by their $\hat{S}^z$ eigenvalues $m_s \in \{\pm \frac{3}{2},\pm \frac{1}{2}\}$. 
The particle-hole spinor is ``locked'' to the physical spin, as it points along the $+ \sigma^1$ ($- \sigma^1$)
direction for positive (negative) $m_s$. The form of the envelope function $f_{m_s}(z)$
is appropriate for the weak pairing limit, $(3 \Delta_p / 4)^2  < \mu$. 

To obtain the effective surface Hamiltonian for nonzero transverse momentum, we diagonalize 
$\hat{h}_1$ in the basis of zero modes given by Eq.~(\ref{psi0-ZM}). 
The only non-vanishing elements obtain from 
\begin{align}
	\hat{h}_1
	\rightarrow&\,
	\sigh^2
	\frac{\Delta_p}{2}
	\left(
	\Sh^+ \kc 
	+
	\Sh^- \kb
	\right)
	-
	\sigh^3
	\frac{\gamma}{4}
	\left[
	(\Sh^+)^2 \kc^2 
	+
	(\Sh^-)^2 \kb^2 
	\right].
\end{align}
The first term connects the $\pm 1/2$ states, giving the $\{\kc,\kb\}$-linear terms
in Eq.~(\ref{hsDef}). The second term mixes the $\{3/2,-1/2\}$ and $\{1/2,-3/2\}$ states,
giving the $\{\kc^2,\kb^2\}$ terms in Eq.~(\ref{hsDef}). The parameter
$c \propto - \gamma / \Delta_p$.

\subsubsection{Vanishing SOC in the bulk}

If the SOC parameter $\gamma = 0$, then the $\pm 3/2$ surface bands remain flat in 
degenerate perturbation theory. Another way to get nonzero $c$ in Eq.~(\ref{hsDef}) 
is by incorporating an additional weak $s$-wave pairing amplitude $\Delta_s$,
as in Eq.~(\ref{h0Def}).
Then the zero energy surface eigenstates with vanishing transverse momentum 
again take the form shown in Eq.~(\ref{psi_0}), but
with the modified envelope function  
\begin{align}\label{fms-SP}
	f_{m_s}(z)
	=&\,
	\frac{1}{\sqrt{\mathcal{N}_{m_s}}}
	\exp\left(-\frac{\Delta_p |m_s| z}{2}\right)
	\,
	\sinh\left[z \sqrt{\frac{\Delta_p^2 m_s^2}{4} - \mu - i \Delta_s \sgn(m_s)}\right].
\end{align}
Here we assume that $0 < \Delta_s \ll \Delta_p^2$ (so as to remain in the bulk topological phase with winding number
$\nu= 4$ \cite{Wu2016}). For convenience, we also assume intermediate strength pairing such that
$0 < \mu < (\Delta_p/4)^2$; in this case there is only one branch of bulk scattering states. 

To obtain the effective surface Hamiltonian for nonzero transverse momentum, we use k$\cdot$p theory.
The matrix elements of $\hat{h}_1$ [Eq.~(\ref{h1Def}) with $\gamma = 0$] give the $\{\kc,\kb\}$-linear terms 
in Eq.~(\ref{hsDef}), which connect the $m_s = \pm 1/2$ states. To connect the $\pm 3/2$ states to the former,
one has to go to second order. This yields the matrix elements
\cite{Sakurai}
\begin{align}\label{MatElem}
	-
	\bra{\psi_{0,m_s}}
	\hat{h}_1(\vex{k})
	\,
	\hat{P}
	\,
	\hat{h}_0^{-1}
	\,
	\hat{P}
	\,
	\hat{h}_1(\vex{k})
	\ket{\psi_{0,m_s'}},
\end{align}
where $\hat{P}$ projects out of the degenerate zero mode space. 
Eq.~(\ref{MatElem})
can be expressed using the basis of bulk scattering
states with zero transverse momentum:
\begin{align}\label{kp2nd}
	\bra{\psi_{0,m_s}}
	\hat{h}_1(\vex{k})
	\,
	\hat{P}
	\,
	\hat{h}_0^{-1}
	\,
	\hat{P}
	\,
	\hat{h}_1(\vex{k})
	\ket{\psi_{0,m_s'}}
	=
	\sum_{m_s''}
	\int_0^\infty
	\frac{d q}{\e_{m_s''}(q)}
	\left[
	\begin{aligned}
	&\,
	\bra{\psi_{0,m_s}}\hat{h}_1(\vex{k})\ket{\psi_{q,m_s''}}
	\bra{\psi_{q,m_s''}} \hat{h}_1(\vex{k})\ket{\psi_{0,m_s'}}
	\\&\,
	+
	\bra{\psi_{0,m_s}}\hat{h}_1(\vex{k}) \, \sigh^2 \hat{R} \ket{\psi_{q,m_s''}^*}
	\bra{\psi_{q,m_s''}^*} \sigh^2 \hat{R} \, \hat{h}_1(\vex{k})\ket{\psi_{0,m_s'}}
	\end{aligned}
	\right],
\end{align}
where $\ket{\psi_{q,m_s''}}$ denotes a scattering state with standing wave momentum $q$
(oscillation in the $z$-direction), $\hat{S}^z$-eigenvalue $m_s''$, 
and gapped positive energy eigenvalue $\e_{m_s''}(q)$,
while 
$\sigh^2 \hat{R} \ket{\psi_{q,m_s''}^*}$ is the (negative energy) particle-hole conjugate of $\ket{\psi_{q,m_s''}}$.
The matrix $\hat{R}$ was introduced in Eq.~(\ref{RDef}).  

The scattering states take the form 
\begin{align}\label{psiq}
	\psi_{q,m_s}(z)
	=
	\frac{1}{\sqrt{\mathcal{N}_{q,m_s}}}
	\left\{
	\hat{\alpha}_{q,m_s}
	\left[
	\cos(q z) 
	-
	e^{- \lambda_{q,m_s} z}
	\right]
	+
	\hat{\beta}_{q,m_s}
	\,
	\sin(q z)
	\right\}
	\ket{m_s},
\end{align}
where $\hat{\alpha}_{q,m_s}$ and $\hat{\beta}_{q,m_s}$ are 2-component
spinors in particle-hole space. The expressions for these and 
$\lambda_{q,m_s}$ are unwieldy so we omit them here. 

Finally, one computes Eq.~(\ref{kp2nd}) using Eqs.~(\ref{psi_0}), (\ref{fms-SP}), and (\ref{psiq}).
This second-order result \emph{vanishes} for $\Delta_s = 0$. Nonzero $c$ in Eq.~(\ref{hsDef}) is 
symmetry allowed, and thus expected in the generic situation.
The simplest way to get it is by retaining nonzero $\Delta_s$ in the bound states
$\{\ket{\psi_{0,m_s}}\}$, but neglecting it in the scattering states
$\{\ket{\psi_{q,m_s}}\}$ (which become very complicated for $\Delta_s \neq 0$). 
This gives nonzero terms 
mixing the $\{3/2,-1/2\}$ and $\{1/2,-3/2\}$ states
proportional to $\kc^2$ and $\kb^2$ in $\hs$ 
(above and below the diagonal, respectively, consistent with planar rotational invariance).
Without loss of generality, we can take the coefficients to be real and positive since
the phases can be removed with a unitary transformation.


\subsection{Calculation of the surface winding number in Fig.~\ref{Fig--TQHEP}}

We compute the surface winding number $W$ for the clean, noninteracting
Majorana surface fluid perturbed by time-reversal breaking ``mass'' terms. 
The Hamiltonian is 
\begin{align}\label{hm1m2}
	\hat{h}_{m_1,m_2}(\vex{k})
	\equiv
	\hs(\vex{k})
	+
	m_1 \, \tauh^3 
	+ 
	m_2 \, \kaph^3,
\end{align}
where $\hs$ was defined by Eq.~(\ref{hsDef}). 
The energy bands of $\hat{h}_{m_1,m_2}(\vex{k})$ are gapped
for non-zero values of $m_{1,2}$ unless $m_1 = \pm m_2$, in
which case a gapless linear Dirac point appears at $\vex{k} = 0$. 

Since the mass terms break surface time-reversal symmetry [Eq.~(\ref{Chiral-Surf})], 
the surface theory resides in class D \cite{SRFL2008}. 
In 2D, this class can exhibit a thermal quantum Hall effect 
\cite{KaneFisher,ReadGreen2000,Capelli02},
where edge states carry a quantized energy current. 
The thermal Hall conductivity $\kappa_{xy}$ 
can be expressed in terms of a winding number $W$ via 
\cite{Zhang2011,TSCRev2016}
\begin{align}
	\kappa_{xy} = W \, \kappa_\circ,
\end{align}
where $\kappa_\circ$ was defined by Eq.~(\ref{kapcirc}). 
In terms of the surface Green's function
\begin{align}\label{G}
	\hat{G}(\omega,\vex{k},m_1,m_2)
	\equiv
	\left[
	- 
	i \, \omega \, \hat{1}
	+
	\hat{h}_{m_1,m_2}(\vex{k})
	\right]^{-1},
\end{align}
the winding number is given by \cite{Volovik2003}
\begin{align}\label{W}
	W(m_1,m_2)
	\equiv
	\frac{\epsilon_{\alpha \beta \gamma}}{3! (2 \pi)^2}
	\int_{-\infty}^\infty 
	d \omega
	\int_{\mathbb{R}^2}
	d^2\vex{k}
	\,
	\Tr
	\left[
	\left(
	\hat{G}^{-1}
	\partial_\alpha
	\hat{G}
	\right)
	\left(
	\hat{G}^{-1}
	\partial_\beta
	\hat{G}
	\right)
	\left(
	\hat{G}^{-1}
	\partial_\gamma
	\hat{G}
	\right)
	\right],
\end{align}
where $\Tr$ denotes the trace over spin-3/2 components
and $\alpha,\beta,\gamma \in \{\omega,k_x,k_y\}$. 
Numerical evaluation of Eq.~(\ref{W}) using Eqs.~(\ref{hm1m2}) and (\ref{G}) 
leads to the winding number results shown in Fig.~2.


\subsection{Perturbative vertex renormalization}

The imaginary time action for the clean, time-reversal invariant, 
interacting Majorana surface theory implied by Eqs.~(\ref{hsDef}) and (\ref{HisDef}) is given by
\begin{align}\label{SClean}
	S 
	= 
	\frac{1}{2}
	\int
	\frac{d \omega \, d^2 \vex{k}}{(2 \pi)^3}
	\eta^\T(-\omega,-\vex{k})
	\,
	\Mps
	\,
	\left[
	-i \omega 
	+
	\hs(\vex{k})
	\right]
	\eta(\omega,\vex{k})
	+
	\frac{u}{4!}
	\int
	d \tau \, 
	d^2 \vex{r}
	\,
	\epsilon_{i_1 i_2 i_3 i_4}
	\,
	\eta_{i_1}\, \eta_{i_2}\, \eta_{i_3} \, \eta_{i_4},
\end{align}
where we have antisymmetrized the four-fermion interaction using
the fourth-rank Levi-Civita tensor. Repeated indices are summed. 

To one loop, the bare vertex function evaluates to 
\begin{align}\label{Gamma4Def}
	\left(\Gamma^\pup{4}\right)_{i_1 i_2 i_3 i_4}
	=
	-
	u
	\,
	\epsilon_{i_1 i_2 i_3 i_4}
	+
	\frac{u^2}{2}
	&
	\left[
		\epsilon_{i_1 i_2 j_1 j_2} \epsilon_{j_3 j_4 i_3 i_4} 
		+ 
		\epsilon_{i_1 i_3 j_1 j_2} \epsilon_{j_3 j_4 i_4 i_2} 
		+ 
		\epsilon_{i_1 i_4 j_1 j_2} \epsilon_{j_3 j_4 i_2 i_3} 
	\right]
\nonumber\\
	&\,
	\times
	\int
	\frac{d \omega \, d^2 \vex{k}}{(2 \pi)^3}
	\left[\hat{G} \Mps\right]_{j_1 j_3}(\omega,\vex{k})
	\,
	\left[\hat{G} \Mps\right]_{j_4 j_2}(\omega,\vex{k}),
\end{align}
valid in the limit of vanishing external frequencies and momenta. 
The three double Levi-Civita terms in the square brackets correspond to the three loop corrections shown 
in Fig.~\ref{SM:Fig--VC}.
We emphasize that the sign of each diagram has to be carefully 
determined using Wick's theorem for the Majorana fermion field. 
The Green's function $\hat{G}(\omega,\vex{k})$ is given by 
Eq.~(\ref{G}) with $m_1 = m_2 = 0$, while $\Mps$ was defined by 
Eq.~(\ref{PH-Surf}).

	\begin{figure}
	\includegraphics[angle=0,width=.4\textwidth]{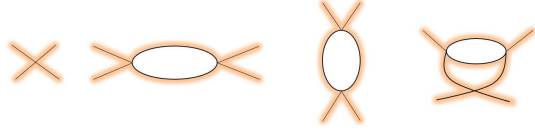}
	\caption{
	Feynman diagrams for the one-loop vertex corrections. 
	}
	\label{SM:Fig--VC}
	\end{figure}

We define
\[
	D(\omega,k) \equiv c^4 k^8+2 c^2 k^4 \omega^2+k^2 \omega^2+\omega^4 
\]
and 
\begin{align}
	\mathcal{N}_{i_1 i_2 i_3 i_4}(\omega,k) 
	\equiv 
	\int_0^{2 \pi} d \phi_k \, 
	D^2(\omega,k) 
	\, 
	\left[\hat{G} \Mps\right]_{i_1 i_2}(\omega,k,\phi_k) 
	\, 
	\left[\hat{G} \Mps\right]_{i_3 i_4}(\omega,k,\phi_k),
\end{align}
where we have switched to polar momentum coordinates $\vex{k} \rightarrow (k,\phi_k)$. 
Next we compute
\begin{multline}
	\frac{1}{2} 
	\left[
	\epsilon_{i_1 i_2 j_1 j_2} \epsilon_{j_3 j_4 i_3 i_4}
	+
	\epsilon_{i_1 i_3 j_1 j_2} \epsilon_{j_3 j_4 i_4 i_2}
	+
	\epsilon_{i_1 i_4 j_1 j_2} \epsilon_{j_3 j_4 i_2 i_3}
	\right]
	\mathcal{N}_{j_1 j_3 j_4 j_2}(\omega \equiv c k^2 x,k)
\\	
	=
	-
	4 	
	c^4 k^{10} 
	\pi
	\left[
		x^2 + c^2 k^2(1 + x^2)^2
	\right]
	\epsilon_{i_1 i_2 i_3 i_4}.
\end{multline}
Thus Eq.~(\ref{Gamma4Def}) reduces to 
\begin{align}\label{Gamma4--2}
	\left(\Gamma^\pup{4}\right)_{i_1 i_2 i_3 i_4}
	=&\,
	-
	\epsilon_{i_1 i_2 i_3 i_4}
	\left\{
	u
	\,
	+
	u^2
	\int_{- \infty}^\infty 
	d x
	\,
	\int_0^\Lambda 
	d k 
	\,
	\frac{(c k^3)4 c^4 k^{10} \pi}{2^3 \pi^3} 	
	\,
	\frac{
	\left[
		x^2 + c^2 k^2(1 + x^2)^2
	\right]}
	{D^2(c k^2 x,k)}
	\right\}
\nonumber\\
	=&\,
	-
	\epsilon_{i_1 i_2 i_3 i_4}
	\left\{
	u
	\,
	+
	\frac{u^2}{4 \pi c}
	\ln\left[\sqrt{(2 c \Lambda)^2 + 1} + 2 c \Lambda \right]
	\right\},
\end{align}
where $\Lambda$ denotes the ultraviolet momentum cutoff. 
Taking the limit $c \Lambda \gg 1$ gives Eq.~(\ref{Vertex}).


\subsection{Mean-field theory: surface thermal quantum Hall plateaux \label{Sec: MFT}}

The interaction strength $u$ is enhanced (suppressed)
by quantum fluctuations for $u > 0$ ($u < 0$) [Eq.~(\ref{uBetaClean})].
In Eq.~(\ref{HisDef}) and the text following, it is noted that $u > 0$ 
is an attractive (repulsive) interaction in the ``$\tauh^3$''
(``$\kaph^3$'') channel, where these matrices specify mass
terms used to construct the thermal quantum Hall phase diagram
shown in Fig.~\ref{Fig--TQHEP}. 
We therefore expect that spontaneous symmetry breaking due
to quantum fluctuations for positive $u$ can be characterized
by an order parameter
\begin{align}\label{MDef}
	\mathcal{M} 
	\equiv 
	\frac{u}{2(1 - 2 \alpha)}
	\left\langle \eta^\T \, \Mps \, \hat{\Lambda}_\alpha \, \eta \right\rangle,
\end{align}
where
\begin{align}
	\hat{\Lambda}_\alpha \equiv (1 - \alpha) \tauh^3 + \alpha \kaph^3
\end{align}
and $\alpha$ is a real variational parameter.
In terms of Majorana (spin-3/2) components,
\begin{align}\label{Mcomp}
	\frac{1}{2}
	\eta^\T \, \Mps \, \hat{\Lambda}_\alpha \, \eta
	= 
	\eta_1 \eta_4 + (1 - 2 \alpha) \, \eta_3 \eta_2.
\end{align}

The interaction in Eq.~(\ref{HisDef}) can be written as
\begin{align}\label{genpair2}
	\His
	=
	-
	\frac{(1 - 2 \alpha)}{2 u}
	\int
	d^2 \vex{r}
	\,
	\left\{
	\mathcal{M} 
	+ 
	\left[
		\frac{u}{2(1 - 2 \alpha)} \eta^\T \, \Mps \, \hat{\Lambda}_\alpha \, \eta
		-
		\mathcal{M}
	\right]
	\right\}^2.
\end{align}
The interaction is attractive for $u > 0$ so long
as $0 \leq \alpha < 1/2$. 
Precisely for $\alpha = 1/2$, Eq.~(\ref{Mcomp}) 
implies that 
$\left(\frac{1}{2}\eta^\T \, \Mps \, \hat{\Lambda}_\alpha \, \eta\right)^2 = 0$ 
due to Pauli exclusion
(neglecting nontrivial anticommutators).
In this case the interaction cannot be written as the square
of the bilinear. 

The zero temperature mean-field condensation energy density is given by 
\begin{align}\label{DeltaE}
	\Delta \mathcal{E}(\mathcal{M})
	=&\,
	(1 - 2 \alpha)
	\frac{\mathcal{M}^2}{u}
	-
	\frac{1}{2 \pi}
	\int_0^{\Lambda} 
	k 
	d k
	\left\{
	\left[\e_1(k,\mathcal{M}) - \e_1(k,0)\right]
	+
	\left[\e_3(k,\mathcal{M}) - \e_3(k,0)\right]
	\right\},
\end{align}
where $\e_1(k,\mathcal{M})$ and $\e_3(k,\mathcal{M})$
denote the linear and cubic surface band energies modified
by the addition of the term $\mathcal{M} \, \hat{\Lambda}_\alpha$ 
to $\hs$ in Eq.~(\ref{hsDef});  
$\e_1(k,0)$ and $\e_3(k,0)$ are the unperturbed, gapless
linear and cubic dispersion relations
[Eq.~(\ref{EBands})].	

Setting $\alpha = 0$ 
(such that $\hat{\Lambda}_\alpha = \tauh^3$)
and extremizing $\Delta \mathcal{E}(\mathcal{M})$
with respect to $\mathcal{M}$ 
leads to the mean-field result in Eq.~(\ref{MFT}), valid in the weak coupling 
limit $u \ll c$. 
Since quantum fluctuations enhance positive $u$ and the interaction is
attractive in the $\hat{\Lambda}_\alpha$ channel only for $0 \leq \alpha < 1/2$,
we expect that the order is weighted towards $\tauh^3$ instead of $\kaph^3$;
the surface winding number $W = + 2$ throughout this range (Fig.~\ref{Fig--TQHEP}). 

A curious aspect of 
Eq.~(\ref{DeltaE}) is the following. By choosing $\alpha$ arbitrarily close
to $1/2$, we can suppress the contribution of the first term on the right-hand side
of Eq.~(\ref{DeltaE}). This allows us to take larger and larger values for $\mathcal{M}$
so as to enhance the negativity of the second term. Yet it cannot be that the system
wants to condense with the bilinear with $\hat{\Lambda}_\alpha = (\tauh^3 + \kaph^3)/2$,
since the interaction cannot even be written in terms of its square (as discussed above).
This suggests that the true admixture of $\tauh^3$ and $\kaph^3$ order [i.e.,
the value of $\alpha$ in the expectation value of $\mathcal{M}$ defined by Eq.~(\ref{MDef})]
cannot be determined by mean-field theory. This warrants further investigation, but 
we will not pursue it here.


\section{Quenched disorder \label{Sec: Dirty}}


\subsection{The ``standard'' theory for a disordered class DIII system: thermal metal \label{Sec: DirtyTheory}}

Next we comment on the physics of the noninteracting, disordered Majorana surface fluid.
Since the spin-3/2 model with surface Hamiltonian given by Eq.~(\ref{hsDef}) has a cubic van Hove singularity,
one would naively expect the ``standard program'' \cite{Altland2010} for
deriving the effective low-energy field theory 
in the presence of disorder
would apply. The steps in this program
are 
\begin{enumerate}
\item{Write a (replicated) Grassmann path integral in order to compute disorder-averaged products of retarded 
and advanced Green's functions.}
\item{Average over the disorder potentials $P_{1,2}(\vex{r})$ in Eq.~(\ref{Dirtyhs}) with variance $\lambda$.}
\item{Decouple the four-field term using an unconstrained matrix field $\hat{Q}$.}
\item{Integrate out the Grassmann field.}
\item{Compute the saddle-point configuration of $\hat{Q}$, the strength of which is the 
self-consistent Born approximation for the elastic scattering rate. This should smear out the 
van Hove singularity in the clean density of states.} 
\item{Perform a gradient expansion and integrate out massive modes to get the nonlinear
sigma model for the constrained matrix field $\hat{Q}$ in the appropriate symmetry class. 
Fluctuations due to quantum interference are controlled by the (inverse of the) coupling constant $G$,
which is the dimensionless charge, spin, or thermal longitudinal dc conductance (depending upon the class and context).}
\end{enumerate}

For a gapless, 2D class DIII Majorana system as described here, this program was carried out in a nontopological
context in \cite{SenthilFisher2000}. The resulting theory has the thermal conductance determined by 
the bare strength of the disorder $G \propto 1/\lambda$,
and $G$ grows with increasing system size due to weak antilocalization. Although the global density of states (DoS) diverges
and wave functions are weakly multifractal, neither are universal. 

Our numerical results in Figs.~\ref{Fig--IDOS} and \ref{Fig--MFC} instead imply universal behavior in the disorder-averaged DoS and multifractal
spectrum for the spin-3/2 surface Majorana fluid, consistent with the SO$(n)_4$ conformal field theory. 
Interestingly, the latter has a sigma model description (non-abelian bosonization) 
that is almost identical to the one obtained by the ``standard program,'' but augmented with a
Wess-Zumino-Novikov-Witten term \cite{WZWP4}. Yet the key point is that the coupling strength is pinned to
a special value equal to the winding number 4, times a universal constant. Since the winding number is not large,
the field theory is strongly coupled. This is 
in part
why the standard program fails in this case. 
For the spin-1/2 TSC models studied previously \cite{WZWP2,WZWP4,WZWP3}, a nonperturbative derivation
was possible using conformal embedding theory. This is not possible in the present case, since the clean
surface with Hamiltonian given by Eq.~(\ref{hsDef}) is not a conformal field theory (as evidenced by the fact that $c$ has
units of length). 

Finally, we stress an important technical point. The CFT is an \emph{unstable} fixed point of
the sigma model. In the absence of interactions, the sigma model is characterized by a single coupling
strength $\lambda$, which can be understood as the dimensionless thermal resistance of the system. 
In the large winding number limit $|\nu| \gg 1$, it is possible to compute the beta function for
$\lambda$, incorporating the WZNW term. The result is \cite{WZWP3}
\begin{align}\label{DIIIBeta}
	\frac{d \lambda}{d l}
	=
	-
	2 \lambda^2
	\left[
	1
	-
	(|\nu| \lambda)^2
	\right],
\end{align}
valid for $\lambda \leq |\nu|$. 
The CFT has $\lambda = 1 / |\nu|$, and this remains a fixed point to all orders in $\lambda$. 
However, any $\lambda < 1 / |\nu|$ flows to ever smaller $\lambda$; this is the signal
for weak antilocalization in the thermal metallic phase \cite{SenthilFisher2000}. 
The fact that our numerical results for the spin-3/2 Majorana surface fluid studied in this
work coincide with the SO$(n)_4$ fixed point implies that the topology ``fine tunes'' $\lambda$
to its fixed point value. Since $\lambda$ is an inverse conductance, it implies
the quantization of the thermal conductivity [Eq.~(\ref{QuantKappa})], which remains
true even when interaction (Altshuler-Aronov) corrections are taken into account \cite{WZWP3}.


\subsection{Momentum space exact diagonalization \label{Sec: Num}}

We consider the effective surface Hamiltonian in the presence of the most general 
form of time-reversal invariant disorder, as given in Eq.~(\ref{Dirtyhs}). 
For a system of finite size $L$, the Hamiltonian can be expressed as a sum over 
discrete points in momentum space:
\begin{gather}\label{msum}
	H 
	= 
	\frac{1}{2}
	\int d^2\vex{r}\, 
	\eta^\T(\vex{r}) \,
	\Mps 
	\left[
		\hs
		+
		\hat{\vex{\tau}}\cdot\vex{P}(\vex{r})
	\right]
	\eta(\vex{r}) 
	= 
	\frac{1}{2}
	\sum_{\vex{n},\vex{m}} 
	\eta^T_{-\vex{m}}\,
	\Mps
	\,
	[\hsP]_{\vex{m},\vex{n}}
	\,
	\eta_{\vex{n}},
\nonumber\\
	[\hsP]_{\vex{m},\vex{n}}
	\equiv
		\delta_{\vex{m},\vex{n}}
		\,
		\hs\!\left(\vex{k} = {\ts{\frac{2 \pi}{L}}} \vex{n}\right)
		+
		\hat{\vex{\tau}}\cdot\vex{P}_{(\vex{m}-\vex{n})},
\end{gather}
with the Fourier conventions
\[
	\eta_{\vex{n}} 
	= 
	\frac{1}{L}\int d^2\vex{r}\ e^{-i\frac{2\pi}{L}\vex{n}\cdot\vex{r}}\ \eta(\vex{r}),
\;\;
	\eta(\vex{r}) 
	= 
	\frac{1}{L}\sum_{\vex{n}}\  e^{i\frac{2\pi}{L}\vex{n}\cdot\vex{r}}\ \eta_{\vex{n}},
\;\;
	P_{\mu,\vex{n}} 
	= 
	\frac{1}{L^2}\int d^2\vex{r}\ e^{-i\frac{2\pi}{L}\vex{n}\cdot\vex{r}}\ P_{\mu}(\vex{r}),
\;\;
	P_{\mu}(\vex{r}) 
	= 
	\sum_{\vex{n}}\  e^{i\frac{2\pi}{L}\vex{n}\cdot\vex{r}}\ P_{\mu,\vex{n}}.
\]
In these equations $\vex{n} \in \{\mathbb{Z},\mathbb{Z}\}$ and the components of
$P_\mu$ are $\mu \in \{1,2\}$. 

For exact diagonalization, we choose a momentum cutoff $N_k$ and keep only those points 
$\vex{n}$ in momentum space with 
$	
	-N_k \leq n_i \leq N_k$, 
for 
$i = 1,2$. 
This choice corresponds to an energy cutoff of 
$\Lambda = 2\pi N_k/L.$ 
We also approximate the Gaussian white noise disorder potentials $P_{\mu,\vex{n}}$ with random-phase Gaussian amplitude distributions via 
\begin{align}\label{Gauss}
	P_{\mu,\vex{n}} 
	\Rightarrow 
	\frac{\sqrt{\lambda}}{L}
	\exp\left(-\frac{\pi^2}{L^2}\xi^2\vex{n}^2 + i \, \theta_{\mu,\vex{n}}\right),
\end{align}
where $\lambda$ is the dimensionless variance of the disorder, 
$\xi$ is a short-distance correlation length of the order $L / N_k$, 
and $\theta_{\mu,\vex{n}} \in [0,2\pi)$ is a uniformly distributed random phase angle. 
Since $P_{\mu}(\vex{r})$ is a real-valued disorder potential, the phases are taken 
to satisfy
$ \theta_{\mu,\vex{n}} = - \theta_{\mu,-\vex{n}}$.
The random-phase approach is equivalent to the disorder-average up to finite-size corrections \cite{YZCP1}. 
The resulting approximate Hamiltonian 
$[\hsP]_{\vex{m},\vex{n}}$
is a dense numerical matrix acting on a 
$4(2 N_k + 1)^2$-dimensional 
Hilbert space that we diagonalize to obtain the energy values and eigenstate wave functions. 

In the plots shown in Figs.~3 and 4, we set 
$\xi = 0.25 \, (L/N_k)$ 
and 
$c = 0.078 \, (L/N_k)$ (such that $2 \pi N_k c / L = 0.49$ for $N_k = 40$). 
Our results are robust with respect to variation of the system size $N_k$, the correlation length $\xi$, and $c$. 
In our calculations, we have retained $P_{\mu,0}$ given by Eq.~(\ref{Gauss}) with  $\theta_{\mu,0} = 0$. 
This represents a nonzero average disorder strength proportional to $\sqrt{\lambda}$. The associated matrix
bilinears $\tauh^{1,2}$ break rotational invariance. Thus our numerical results show good agreement with the 
CFT even when we incorporate nonzero (but weak) average anisotropy. We have also performed calculations with 
$P_{\mu,0} = 0$. The results are indistinguishable.


\section*{Acknowledgements}

We thank Bitan Roy for helpful discussions.
This research was supported by the Welch Foundation grants No.~E-1146 (S.A.A.G.) and No.~C-1809 
(S.D. and M.S.F.), as well as NSF CAREER grant No.~DMR-1552327 (S.D. and M.S.F.),
and by the Texas center for the superconductivity (S.A.A.G.).

\end{widetext}


\begin{thebibliography}{100}
\bibitem{TIRev2010}  
	M. Z. Hasan and C. L. Kane, Rev. Mod. Phys. \textbf{82}, 3045 (2010); 
	X.-L. Qi and S.-C. Zhang, \textit{ibid.} \textbf{83}, 1057 (2011).
\bibitem{BernevigHughes2013}
	B. A. Bernevig and T. L. Hughes,
	\textit{Topological Insulators and Topological Superconductors}
	(Princeton University Press, Princeton, New Jersey, 2013). 
\bibitem{TSCRev2016} 
	C.-K. Chiu, J. C. Y. Teo, A. P. Schnyder, and S. Ryu, Rev. Mod. Phys. {\bf 88}, 035005 (2016).
\bibitem{TSCRev2016B} 
	T. Mizushima, Y. Tsutsumi, T. Kawakami, M. Sato, M. Ichioka, and K. Machida,	
	J. Phys. Soc. Jpn. {\bf 85}, 022001 (2016).
\bibitem{Volovik2003} 
	G.E. Volovik, \textit{The Universe in a Helium Droplet} (Oxford University Press, Oxford, 2003).
\bibitem{SRFL2008} 
	A. P. Schnyder, S. Ryu, A. Furusaki, and A. W. W. Ludwig, Phys. Rev. B \textbf{78}, 195125 (2008).
\bibitem{WZWP2} 
	M. S. Foster and E. A. Yuzbashyan, Phys. Rev. Lett. \textbf{109}, 246801 (2012).
\bibitem{WZWP4} 
	M. S. Foster, H.-Y. Xie, and Y.-Z. Chou, Phys. Rev. B \textbf{89}, 155140 (2014).
\bibitem{YZCP1}
	Y.-Z. Chou and M. S. Foster, Phys. Rev. B {\bf 89}, 165136 (2014). 
\bibitem{WZWP3} 
	H.-Y. Xie, Y.-Z. Chou, and M. S. Foster, Phys. Rev. B \textbf{91}, 024203 (2015).
\bibitem{Wu2016} 
	W. Yang, Y. Li, and C. Wu, 
	Phys. Rev. Lett. \textbf{117}, 075301 (2016).
\bibitem{Fang2015}
	C. Fang, B. A. Bernevig, and M. J. Gilbert,
	Phys. Rev. B {\bf 91}, 165421 (2015).
\bibitem{Paglione2016}
	H. Kim, 
	K. Wang, 
	Y. Nakajima, 
	R. Hu, 
	S. Ziemak, 
	P. Syers, 
	L. Wang, 
	H. Hodovanets, 
	J. D. Denlinger, 
	P. M. R. Brydon, 
	D. F. Agterberg, 
	M. A. Tanatar, 
	R. Prozorov, 
	and 
	J. Paglione,
	arXiv:1603.03375.
\bibitem{Brydon2016}
	P. M. R. Brydon, L. Wang, M. Weinert, and D. F. Agterberg,
	Phys. Rev. Lett. {\bf 116}, 177001 (2016).
\bibitem{KaneFisher}
	C. L. Kane and M. P. A. Fisher,
	Phys. Rev. B {\bf 55}, 15832 (1997).
\bibitem{ReadGreen2000}
	N. Read and D. Green,
	Phys. Rev. B {\bf 61}, 10267 (2000).
\bibitem{Capelli02}
	A. Capelli, M. Huerta, and G. Zemba, 
	Nucl. Phys. B {\bf 636}, 568 (2002).
\bibitem{Zhang2011}
	Z. Wang, X.-L. Qi, and S.-C. Zhang,
	Phys. Rev. B {\bf 84}, 014527 (2011). 
\bibitem{Ryu2012}
	S. Ryu, J. E. Moore, and A. W. W. Ludwig,
	Phys. Rev. B {\bf 85}, 045104 (2012). 
\bibitem{Stone2012}
	M. Stone,
	Phys. Rev. B {\bf 85}, 184503 (2012).
\bibitem{SenthilFisher2000}
	T. Senthil and M. P. A. Fisher,
	Phys. Rev. B {\bf 61}, 9690 (2000). 
\bibitem{YuCardona}
	P. Y. Yu and M. Cardona, 
	\textit{Fundamentals of Semiconductors}
	(Springer-Verlag, Berlin, 1996). 
\bibitem{Ghorashi2017}
	S. A. A. Ghorashi, B. Roy, and M. S. Foster, 
	unpublished. 
\bibitem{Nomura2015}
	Y. Shimizu, A. Yamakage, and K. Nomura, 
	Phys. Rev. B {\bf 91}, 195139 (2015).
\bibitem{AndRev08} 
	F. Evers and A. D. Mirlin, Rev. Mod. Phys. \textbf{80}, 1355 (2008).
\bibitem{Chamon96} 
	C. C. Chamon, C. Mudry, and X.-G. Wen, Phys. Rev. Lett. \textbf{77}, 4194 (1996).
\bibitem{MFCP1}
	M. S. Foster, S. Ryu, and A. W. W. Ludwig, 
	Phys. Rev. B {\bf 80}, 075101 (2009);
	T. Vojta, Physics {\bf 2}, 66 (2009).
\bibitem{Feigelman07}
	T. I. Baturina \textit{et al.},
	Physica C {\bf 468}, 316 (2008);
	M. V. Feigel'man, L. B. Ioffe, V. E. Kravtsov, and E. A. Yuzbashyan,
	Phys. Rev. Lett. {\bf 98} 027001 (2007).
\bibitem{Feigelman10}
	M. V. Feigel'man, L. B. Ioffe, V. E. Kravtsov, and E. Cuevas,
	Ann. Phys. {\bf 325}, 1390 (2010).
\bibitem{Maciejko15}
	Y. J. Park, S. B. Chung, and J. Maciejko,
	Phys. Rev. B {\bf 91}, 054507 (2015).
\bibitem{Sakurai}
	J. J. Sakurai,
	\textit{Modern Quantum Mechanics}
	(Addison-Wesley, Reading, Mass., 1994).
\bibitem{Altland2010}
	A. Altland and B. Simons,
	\textit{Condensed Matter Field Theory}, 2nd.\ Ed.\ 
	(Cambridge University Press, Cambridge, England, 2010). 
\end{thebibliography}
\end{document}